\documentclass[numberedappendix,twocolumn,twocolappendix]{openjournal}

\usepackage{latexsym}
\usepackage{graphicx}
\usepackage{amssymb}
\usepackage{longtable}
\usepackage{epsf}
\usepackage{amsmath}
\usepackage{graphicx}
\usepackage{gensymb}

\usepackage[backref=page]{hyperref}
\hypersetup{colorlinks=true,linkcolor=blue,citecolor=blue,filecolor=blue,urlcolor=blue}
\usepackage{cleveref}
\usepackage{bm}

\usepackage{natbib}

\newcommand{\beq}{\begin{eqnarray}}
\newcommand{\eeq}{\end{eqnarray}}
\newcommand{\ben}{\begin{enumerate}}
\newcommand{\een}{\end{enumerate}}
\newcommand{\bit}{\begin{itemize}}
\newcommand{\eit}{\end{itemize}}

\newcommand{\zobs}{z_{\rm obs}}
\newcommand{\Mstar}{M_{\star}}
\newcommand{\tauage}{\tau_{\rm age}}
\newcommand{\thetagal}{\theta_{\rm gal}}

\newcommand{\LSSP}{L_{\rm SSP}}

\usepackage[table]{xcolor} % enables custom text colors
\definecolor{hpurple}{HTML}{7E16DF}
\definecolor{horange}{HTML}{FFA500}
\definecolor{hred}{HTML}{d62728}

\usepackage{xspace}

\begin{document}

\title{Bayesian Posteriors with Stellar Population Synthesis on GPUs}
\shorttitle{SPS Inference on GPUs}

\author{Georgios Zacharegkas$^{1,\star}$}
\author{Andrew Hearin$^{1}$}
\author{Andrew Benson$^{2}$}

\affiliation{$^1$HEP Division, Argonne National Laboratory, 9700 South Cass Avenue, Lemont, IL 60439, USA}
\affiliation{$^2$Carnegie Observatories, 813 Santa Barbara Street, Pasadena, CA 91101}
\thanks{$^{\star}$E-mail:gzacharegkas@anl.gov}

\shortauthors{Zacharegkas, Hearin \& Benson}

%%%%%%%%%%%%%%%%%%%%%%%%%%%%%%%%%%%%%%%%%%%%%%%%%%
\begin{abstract}
Models of Stellar Population Synthesis (SPS) provide a predictive framework for the spectral energy distribution (SED) of a galaxy. SPS predictions can be computationally intensive, creating a bottleneck for attempts to infer the physical properties of large populations of individual galaxies from their SEDs and photometry; these computational challenges are especially daunting for near-future cosmology surveys that will measure the photometry of billions of galaxies. In this paper, we explore a range of computational techniques aimed at accelerating SPS predictions of galaxy photometry using the JAX library to target GPUs. We study a particularly advantageous approximation to the calculation of galaxy photometry that speeds up the computation by a factor of 50 relative to the exact calculation. We introduce a novel technique for incorporating burstiness into models of galaxy star formation history that captures very short-timescale fluctuations with negligible increase in computation time. We study the performance of Hamiltonian Monte Carlo (HMC) algorithms in which individual chains are parallelized across independent GPU threads, finding that our pipeline can carry out Bayesian inference at a rate of approximately $1000$ galaxy posteriors per minute on a single GPU. Our results provide an update to standard benchmarks in the literature on the computational demands of SPS inference; our publicly available code enables previously-impractical Bayesian analyses of large galaxy samples, and includes several standalone modules that could be adopted to speedup existing SPS pipelines.
\end{abstract}


%%%%%%%%%%%%%%%%%%%%%%%%%%%%%%%%%%%%%%%%%%%%%%%%%%
\maketitle

\vspace{1cm}

\twocolumngrid 

%%%%%%%%%%%%%%%%%%%%%%%%%%%%%%%%%%%%%%%%%%%%%%%%%%
%%%%%%%%%%%%%%%%%%%%%%%%%%%%%%%%%%%%%%%%%%%%%%%%%%
\section{Introduction}
\label{sec:intro}

The prevailing theoretical framework for predicting the spectral energy distribution (SED) of a galaxy from its fundamental physical properties is  stellar population synthesis \citep[SPS,][]{conroy13_sps_aa_review}. The concepts and defining equations of SPS have been in place for decades \citep[e.g.,][]{Tinsley1978_sps, Bruzual1983_sps, Arimoto1987_sps, Buzzoni1989_sps,bruzual_charlot_1993,worthey_1994_age_metal_degeneracy,maraston_1998,Leitherer_starburst99}, and there are numerous SPS libraries that are publicly available \citep[e.g.,][]{fioc_rocca_volmerange99_pegase,Bruzual2003,leborgne_etal04,maraston05,conroy_gunn_fsps1,eldridge_etal17,larry_bradley_2020_photutils_v1p0,johnson_etal21_prospector}.

One of the most common forms of SPS analysis is fitting the observed SED or photometry of a galaxy with a physical model \citep[e.g.,][]{sawicki_yee_1998,brinchmann_ellis_2000,salim_etal07_sdss_sfrs,kriek_etal09}. Monte Carlo Markov Chains (MCMCs) are the standard technique for Bayesian uncertainty quantification in such fits, and SPS inference with MCMC can be computationally costly. As a case study for the computational demands of SPS inference, consider \citet{leja_3dhst_2019a}, which used a 14-parameter SPS model implemented in Prospector \citep{johnson_etal21_prospector} to analyze 58,461 galaxies from the 3D-HST survey \citep{momcheva_etal16_3dhst}; the authors report using $\sim10^5-10^6$ likelihood evaluations for each galaxy's MCMC chain to converge, and an average runtime of 25 CPU hours per galaxy, resulting in a total of 1.5 million CPU hours for the analysis.\footnote{Comparable per-galaxy MCMC runtimes with Prospector are also reported in \citet{tacchella_etal22}.} For cosmological datasets with $10^7-10^9$ galaxies, naively scaling traditional SPS inference techniques would require hundreds of millions to billions of CPU-hours; these demands are especially daunting considering that obtaining robust results commonly requires many MCMCs per galaxy, for example when running tests for the effect of different spectral libraries in the analysis.
 
There are essentially two classes of approaches to addressing this computational challenge: speed up the computation of the SPS prediction for the likelihood, or use some alternative to running a traditional MCMC for each galaxy in the sample. Using an emulator of the SPS predictions is a natural way to speed up the cost of evaluating the likelihood. In \citet{hahn_etal23_PROVABGS}, a neural network emulator of their PROVABGS model was used to derive posteriors on galaxies in DESI BGS \citep{hahn_etal23_desi_bgs}. The authors found that their SPS model evaluates in 340 ms compared to 2.9 ms with the emulator, and that with their emulator they achieve MCMC runtimes of $\sim5$ minutes per galaxy. Large performance gains from neural emulation are also reported in \citet{alsing_etal20_speculator}, in which the authors use a training set of 2 million synthetic galaxy SEDs and photometry to train an emulator of SPS predictions by Prospector, which they call Speculator, finding speedup factors of $10^3-10^4$.

As an example of the second class of approach, in \citet{hahn_melchior_2022_accel_bayes_sed} the authors developed a neural density estimator of individual galaxy posteriors, SEDFlow. To train SEDflow, they use the PROVABGS model to generate MCMC posteriors of $\sim10^6$ synthetic galaxies; once trained, SEDflow can infer the posterior of each new galaxy in $\sim1$ CPU-second. When emulating per-galaxy posteriors, the computational demands become driven by size of the training data required to train an accurate network, rather than by the size of the observational dataset.

An alternative, but closely related, approach to per-galaxy MCMCs is {\em population-level modeling}, which adopts a different strategy to analyze the galaxy population. In population-level modeling, some parametric model (typically a neural network) is used to describe the PDF of physical galaxy properties, and the parameters of the model are tuned so that the galaxy population generated by the PDF presents summary statistics that are in close agreement with the corresponding observations. In \citet{li_etal24_popsed}, the authors use a population-level model, POPSED, to recover various summary statistics of the population of galaxies observed in the GAMA survey \citep{driver_etal11_gama_survey}. Another population-level model, Pop-Cosmos \citep{alsing_etal24_pop_cosmos}, was used to capture the statistical distribution of galaxy properties modeled by Prospector, and thereby place constraints on several key scaling relations of the physics of galaxies in the COSMOS-20 dataset. In subsequent studies, in \cite{Thorp_etal24_popcosmos,Thorp_etal25_popcosmos} Pop-Cosmos was used to estimate properties and redshifts for individual galaxies. The authors quote runtimes of 15 GPU-second per galaxy using the pop-cosmos prior, and 0.6 GPU-second per galaxy under the Prospector prior, a significant milestone in the development of fast and accurate SPS parameter posterior inference.

In the present work, we study three computational techniques for improving the efficiency of SPS inference. First, we explore a method for approximating the effect of dust attenuation on the photometry calculation. Second, we evaluate the computational gains derived from evaluating the likelihood on a GPU using the DSPS library \citep{Hearin2023DSPS}. Finally, we study the computational benefits of leveraging gradient information in the inference algorithm. Broadly, we find that with these techniques used in tandem, we reduce the computation time to derive per-galaxy posteriors by factors of 100-1000 relative to running conventional MCMCs on a CPU.

This paper is organized as follows. In Section~\ref{sec_TheoreticalModel} we give an overview of the baseline SPS model we use in our analysis. The present work is primarily concerned with computational methods of SPS, and our primary results are broadly applicable to a wide range of SPS models, and so Section~\ref{sec_TheoreticalModel} contains only a high-level overview of the particular model we use in our results; a detailed description of the SPS model appears in Appendix~\ref{appendix_TheoreticalModel}. We outline our method for speeding up photometry computations by approximating the effect of dust in Section~\ref{sec_Photometry}; additional details about this technique appear in Appendix~\ref{appendix_ApproxPhotoModel}. In Section~\ref{section_FittingPipeline}, we describe our pipeline for deriving Bayesian posteriors for large galaxy samples, and compare the computational performance of our pipeline to standard techniques in the literature. We conclude in Section~\ref{sec_conclusion} with a summary of our results.

%%%%%%%%%%%%%%%%%%%%%%%%%%%%%%%%%%%%%%%%%%%%%%%%%%
\section{SPS model overview}
\label{sec_TheoreticalModel}

In this section, we give a high-level overview of the model for stellar population synthesis (SPS) that we use throughout the paper. Our primary aim in the present work is to reassess the computational demands of deriving Bayesian posteriors on SPS parameters, and our basic conclusions are applicable to a wide range of SPS models, so we relegate a detailed discussion of our particular model to Appendix~\ref{appendix_TheoreticalModel}.

In SPS, the composite spectral energy distribution (SED) of a galaxy observed at redshift $z$ described by a set of parameters, $\theta_{\rm gal},$ is calculated as the weighted sum of template SEDs of Simple Stellar Populations (SSPs); an SSP is defined as a population of stars that formed simultaneously from a homogeneous gas cloud from some assumed initial mass function (IMF); by definition, all the stars in an SSP have the same age, $\tauage,$ and metallicity, $Z$. The composite galaxy SED is then written as:
\begin{align}
\label{eq_galSEDdef}
    L_{\rm gal} (\lambda\vert z,\theta_{\rm gal}) = \Mstar&\sum_{i,j} \LSSP(\lambda|\tauage^i, Z^j) F_{\rm att}(\lambda|\theta_{\rm gal}) \nonumber \\
    &\times P_{\rm SSP}(\tauage^i, Z^j\vert z,\theta_{\rm gal}) \; .
\end{align}
In Equation~\eqref{eq_galSEDdef}, $\Mstar$ is the total stellar mass formed by redshift $\zobs,$ $F_{\rm att}(\lambda)$ is the attenuation function that encodes the absorption of light by dust within the host galaxy, and $P_{\rm SSP}(\tauage, Z)$ is the probability distribution of stars in the galaxy as a function of age and metallicity. In typical SPS analyses, a particular choice is made in advance for the IMF, isochrones, and stellar SED library, and thereafter the SSP SEDs $ \LSSP(\lambda|\tauage, Z)$, are held fixed throughout the computations; we adopt SSP SEDs from MIST \citep{choi_etal16_mist} throughout this work.

For the computational benchmarking purposes of this paper, we adopt a simple parametrized relation for $P_{\rm SSP}(\tauage, Z),$ in which the age and metallicity weights are separable:
\begin{equation}\label{eq:Pssp_variable_separation}
    P_{\rm SSP}(\tauage, Z) \equiv P_{\rm SSP}(\tauage) \times P_{\rm SSP}(Z) \;.
\end{equation} 
We calculate $P_{\rm SSP}(\tauage)$ by an integral of the the star formation history (SFH), which we model as a double power law, coupled with an additional ingredient for short-timescale fluctuations (i.e., ``burstiness"; see Appendix~\ref{sec_SFHModel} for details). We calculate $P_{\rm SSP}(Z)$ as a lognormal distribution with a median determined by a parametrized mass-metallicity relation (MZR) (see Appendix~\ref{sec_MetallicityModel}). 
We note that comparable runtimes to what we report here can also be achieved with alternative SED models that relax the assumption of separability; for example, the DSPS library includes SED models in which the metallicity distribution function varies with stellar age in accord with the time evolution of the MZR, and runtimes in these models are comparable to the separable model we adopt here for simplicity.

%%%%%%%%%%%%%%%%%%%%%%%%%%%%%%%%%%%%%%%%%%%%%%%%%%
%%%%%%%%%%%%%%%%%%%%%%%%%%%%%%%%%%%%%%%%%%%%%%%%%%
\section{Computing approximate photometry}
\label{sec_Photometry}

One of the most common kinds of measurement used to constrain the physical properties of a galaxy is its broadband photometry, i.e., the flux of the galaxy observed through a collection of filters. In this section, we explore how SPS predictions for photometry can be sped up by leveraging a particular approximation in the calculation of broadband flux.

The photometry of a galaxy through a particular filter, $c_{\rm gal},$ is simply the convolution of its composite SED against the transmission curve of the filter, $T(\lambda):$
\beq
c_{\rm gal}(\theta_{\rm gal},z) \equiv \int d\lambda \; T(\lambda\vert z) L_{\rm gal}(\lambda | \theta_{\rm gal},z) \;,
\eeq
where $T(\lambda\vert z)\equiv T(\lambda/(1+z))$ accounts for the redshifting of the SED.
Plugging in Equation~\eqref{eq_galSEDdef} for $L_{\rm gal}(\lambda)$, we have:
\begin{align}
\label{eq_photo_integral_exact}
    &c_{\rm gal}(\theta_{\rm gal}, z) = \sum_{i,j} P_{\rm SSP} (\tau_{\rm age}^i,Z^j | \theta_{\rm gal},z) \nonumber \\
    & \times \int d\lambda \; T(\lambda\vert z) L_{\rm SSP} (\lambda | 
    \tau_{\rm age}^i,Z^j) F_{\rm att}(\lambda | \theta_{\rm gal}).
\end{align}
The wavelength integration on the right-hand side of Equation~\eqref{eq_photo_integral_exact} turns out to be the most computationally expensive piece of the calculation, particularly when using high-resolution templates for $L_{\rm SSP} (\lambda).$ When running MCMCs to derive posteriors on galaxy properties, one must repeat these expensive integrations over and over for each proposed point in the parameter space $\thetagal.$
However, notice that $F_{\rm att}(\lambda|\theta_{\rm gal})$ is the only quantity in the integrand of Equation~\eqref{eq_photo_integral_exact} that depends on $\thetagal.$ Thus if we make the approximation that $F_{\rm att}(\lambda)$ is approximately constant across the support of the transmission curve, then the wavelength integrations could be done in advance of the MCMC, eliminating one of the dominant bottlenecks in the calculation of the likelihood. In the remainder of this section, we quantify the computational benefits and loss of accuracy associated with this approximation.

We make the approximation that the attenuation curve is constant over the range of wavelength spanned by the transmission curve. We define the \textit{effective wavelength} of the filter, $\lambda_{\rm eff}$, as follows:
\begin{align}
\label{eq_lambda_eff}
    \lambda_{\rm eff}(z) \equiv \frac{\int d\lambda \; \lambda \cdot T(\lambda|z)}{\int d\lambda \; T(\lambda|z)} \; .
\end{align}
We now rewrite Equation~\eqref{eq_photo_integral_exact} after pulling the attenuation curve outside of the integral:
\begin{align}
\label{eq_photo_integral_approx}
    &c_{\rm gal}(z,\theta_{\rm gal}) \approx F_{\rm att}(\lambda_{\rm eff}|\theta_{\rm gal}) \\
    &\qquad \times \sum_{i,j} P_{\rm SSP} (\tau_{\rm age}^i,Z^j | \theta_{\rm gal})\cdot c_{\rm SSP}(\tau_{\rm age}^i,Z^j|z) \nonumber
\end{align}
where the quantity $c_{\rm SSP}$ is the broadband flux of the grid of SSP SEDs:
\begin{align}
    c_{\rm SSP}(\tau_{\rm gal}^i,Z^j|z) \equiv \int d\lambda \; T(\lambda\vert z) L_{\rm SSP} (\lambda \vert \tau_{\rm age}^i,Z^j) \; .
\end{align}
The quantities $c_{\rm SSP}$ can be calculated in advance of the analysis at each point on the SSP grid, either at a fixed $z$ if the redshift is known, or on a lookup table, so that calculating the photometry of a model galaxy only requires computing $P_{\rm SSP} (\tau_{\rm age}^i,Z^j | \theta_{\rm gal}),$ and then taking $P_{\rm SSP}$-weighted sums of the grid of $c_{\rm SSP}.$ Variations on this approximation are used by several SSP libraries, for example in CIGALE \citep{burgarella_etal05_cigale,noll_etal09_cigale,boquien_etal19_cigale} and the {\tt SEDITION} code included in UniverseMachine \citep{behroozi_etal19,behroozi_etal20_umachine_jwst}, and in this paper we provide systematic testing of its accuracy and performance benefits in the context of SPS inference.

\begin{figure}
\centering
\includegraphics[width=\columnwidth]{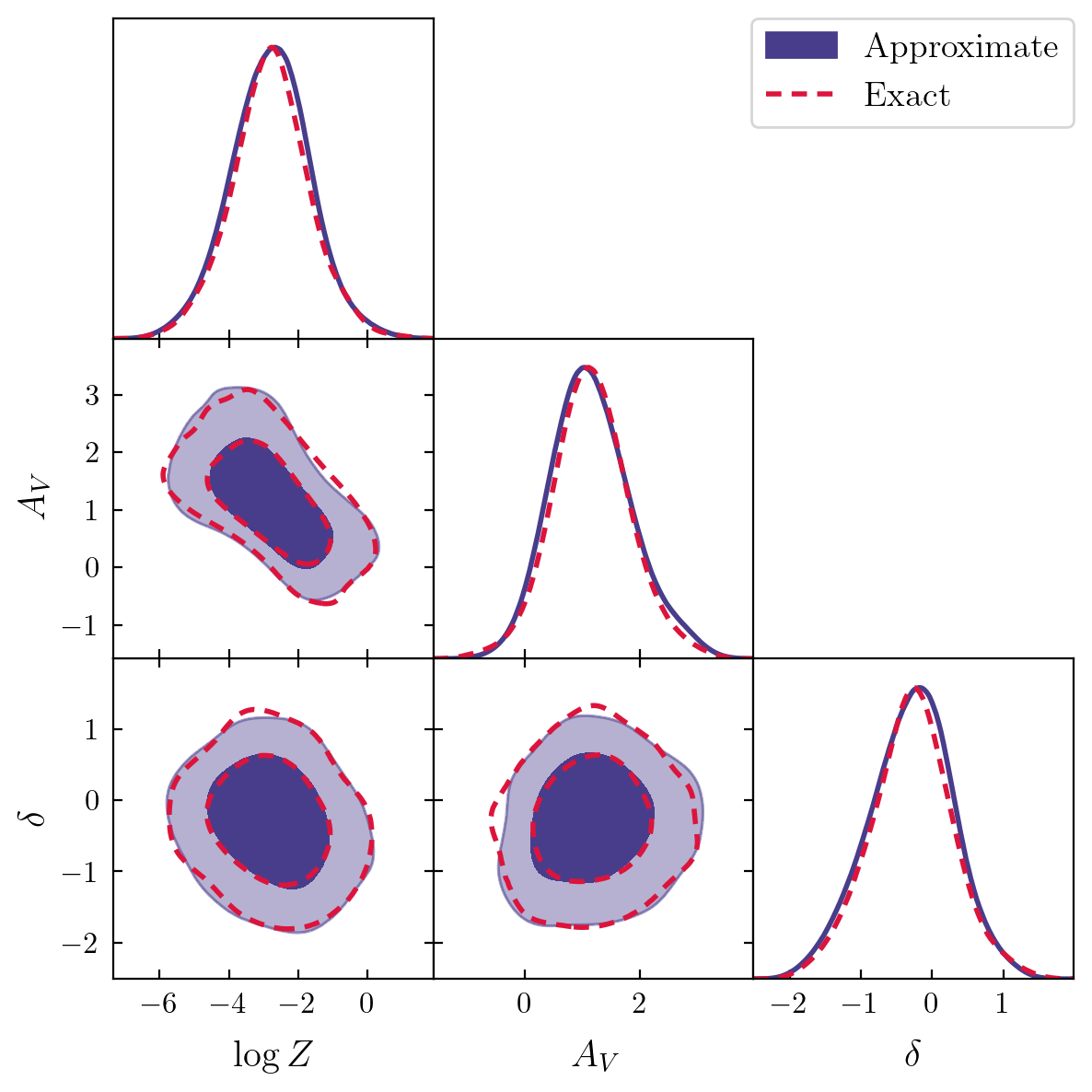}
\caption{\label{fig:ApproxModelTestSomeParamsPartial} Comparison between running a HMC inference using the approximate and exact photometry model, as described in Section~\ref{sec_Photometry}. In this plot we show 3 of the 12 model parameters as a demonstration of how close the two runs look. The full 12-parameter plot is shown in Figure~\ref{appfig:ApproxModelTestAllParams} in Appendix~\ref{appendix_ApproxPhotoModel}.}
\end{figure}

In Appendix \ref{appendix_ApproxPhotoModel}, we present results of two different kinds of tests of the accuracy of the approximation in Equation~\eqref{eq_photo_integral_approx}. First, we have directly checked the level of agreement between the exact and approximate calculation of $c_{\rm gal}$ errors on galaxy magnitudes are limited to $0.1\%$ or better for a wide range of galaxy properties and transmission curves. Second, we have run a full Hamiltonian Monte Carlo (HMC) pipeline to infer the posteriors on an example set of SPS model parameters,  again using the exact vs. approximate photometry; this second test allows us to explicitly propagate the residual errors of the approximation through to our downstream science target. We defer a fuller discussion of our HMC pipeline to Section~\ref{section_FittingPipeline} and Appendix \ref{appendix_ApproxPhotoModel}, but Figure~\ref{fig:ApproxModelTestSomeParamsPartial} summarizes the results of this second test. For the sake of brevity, we plot the posteriors of a subset of 3 model parameters $(\log Z, A_V, \delta),$ derived by running our HMC pipeline until it converges after $3000$ steps. As we can see, Bayesian confidence intervals on the model parameters are nearly indistinguishable for the approximate and exact photometry; as shown in Appendix \ref{appendix_ApproxPhotoModel}, the same is true for the full 12-dimensional parameter space. The HMC run using the exact photometry took approximately 11 hours to complete, whereas the HMC using approximate photometry terminated in 21 minutes, and so providing a speedup factor of over $30$.

%%%%%%%%%%%%%%%%%%%%%%%%%%%%%%%
%%%%%%%%%%%%%%%%%%%%%%%%%%%%%%%

%%%%%%%%%%%%%%%%%%%%%%%%%%%%%%%%%%%%%%%%%%%%%%%%%%
%%%%%%%%%%%%%%%%%%%%%%%%%%%%%%%%%%%%%%%%%%%%%%%%%%

\section{GPU-Parallel SPS Inference}
\label{section_FittingPipeline}

In this section, we study the computational problem of deriving Bayesian posteriors, $\mathcal{P}(\theta_{\rm gal} | D),$ on the parameters $\theta_{\rm gal}$ of an individual galaxy constrained by observations of its photometry, $D.$ As discussed in Section~\ref{sec:intro}, runtimes of $\sim25$ CPU-hour per galaxy are often quoted in the literature for SPS models using variants of MCMC methods to estimate the posterior. Here we revisit this computational benchmark using Hamiltonian Monte Carlo \citep[HMC,][]{duan_etal87_hmc,neal_2011_hmc_chapter}, an alternative Bayesian algorithm that leverages gradient information, and using SPS computations that have been optimized for GPU computing resources. We note that the model we utilized for the benchmarks is not as complex as some SPS models that were employed in the studies whose quoted runtimes we mentioned above. The DSPS library underlying this work is not as full-featured as other, more mature SPS libraries in the field, and in future work we aim to incorporate more realistically complex models into our pipeline. We expect the general trends in computational speedup we present here to continue to hold, since the main reasons for our performance gains are the GPU capability of our code, combined with the ability to leverage gradient-based algorithms, both of which will remain true due to our JAX-based implementation.

Using the SPS model outlined in Section~\ref{sec_TheoreticalModel}, we generate photometry for hundreds of thousands of synthetic galaxies by randomly sampling $\theta_{\rm gal}$ from a broad prior (see Appendix~\ref{appendix:ParameterBounding} for detailed information about the prior distribution we adopted). We treat this photometry as target data, and we run an HMC chain for each synthetic galaxy using the implementation in \href{https://blackjax-devs.github.io/blackjax/}{\texttt{Blackjax}}, a library of samplers written in \texttt{JAX} that can be run on both CPU and GPU machines. We run each chain for a fixed number of steps and check for convergence according to the \textit{Gelman-Rubin} \citep{GelmanRubinRhat} statistic, which compares the variance between and within chains to determine if a run has converged to a stationary state; we adopt $\hat{R} < 1.2$ as a convergence threshold for each chain. 

\begin{figure}
\centering
\includegraphics[width=\columnwidth]{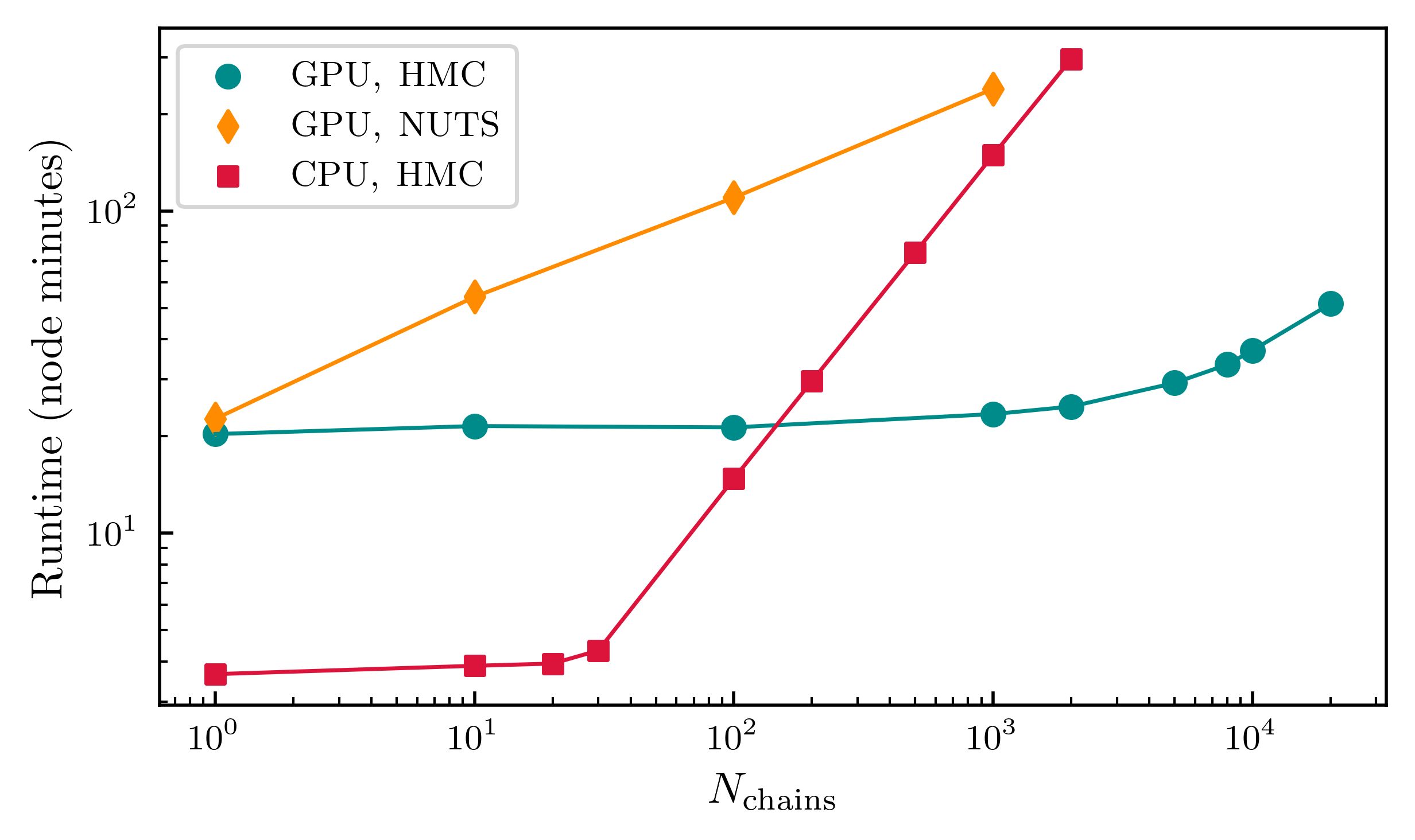}
\caption{\label{fig:GPUvsCPUruntimes} Runtime comparisons between running on a single GPU versus on a CPU as a function of number of fits we perform in parallel. We have run our HMC and NUTS pipelines for $3000$ steps, with $100$ steps of adaptation on both machines, and under the same exact configurations, fitting our full 12-parameter approximate photometry model to the same target data. }
\end{figure}

In Figure~\ref{fig:GPUvsCPUruntimes} we show the runtime for our Bayesian inference pipeline. The horizontal axis shows how many galaxies are in the sample for which posteriors were derived; the vertical axis shows the wall-clock time in node minutes; we show computational performance for three different exercises: HMC run on GPUs and CPUs, as well as an alternative gradient-based posterior sampler implemented in {\tt Blackjax}, the No-U-Turn Sampler \citep[NUTS,][]{hoffman_gelman_2014_nuts}, on GPUs. We run each sampler for $N_{\rm steps}=3000$ steps, with an adaptation phase of $100$ steps, using the exact same target data and sampler settings. We selected $N_{\rm steps}$ to strike a balance between having a high fraction of chains that fully converge, and having a fast runtime. For $N_{\rm steps}=3000$, more than $85\%$ of the fits have fully converged for both HMC and NUTS; the remaining fits are typically close to convergence, and require an additional $\sim N_{\rm steps};$ a small portion of outliers typically requires adjusting the starting point of the chain.

The most striking feature of Figure~\ref{fig:GPUvsCPUruntimes} is the performance of HMC on GPUs: there is virtually no increase in wall-clock time with $N_{\rm chains}\lesssim 2000.$ This reflects both the nature of the GPU architecture as well as the design of our pipeline. GPU computations achieve especially high performance on problems where the same work runs in parallel across many independent GPU threads. For the case of the {\tt Blackjax} implementation of HMC, it is straightforward to parallelize the computation so that each chain is run on an independent GPU thread; since the HMC algorithm does not require iterative calculations in between steps, then each chain requires the same amount of work, and so the chains proceed in lock-step with each other for the pre-determined number of samples. With these considerations in mind, our strategy to optimize these computations is to fit as many independent chains as can fit in the GPU memory; once the available memory is saturated, the runtime begins to rise sharply, which for our benchmark we find occurs for $N_{\rm chains}\gtrsim10^4$. For reference, on the Swing machine we used at Argonne\footnote{https://www.lcrc.anl.gov/systems}, a single GPU has $40$GB of memory; on Amazon Web Services (AWS), one typically uses instances with $16$GB memory, or up to $80$GB for the more powerful systems. Doubling the GPU memory doubles the number of galaxies that can be processed at once on a single GPU. For galaxy samples too large to fit in the memory of a single GPU, it is straightforward to divide the sample into subsets, and process each subset on a different GPU in parallel with MPI.

For the case of a CPU node with $N_{\rm cpu}=30,$ running different HMCs in parallel on independent cores results in nearly constant runtimes for  $N_{\rm chains} < N_{\rm cpu},$ and then increases essentially linearly with $N_{\rm chains}.$ For larger number of chains, $N_{\rm chains}\gtrsim1000,$ the CPU-node calculation is $\sim$15x slower than its GPU-node equivalent. Note also that for $N_{\rm chains} \lesssim 100$ the CPU-node computation is significantly faster, due to the latency of initially transferring data from CPU to GPU, and so the GPU only provides a performance benefit to analyses with large galaxy samples.

The NUTS runtimes for large numbers of galaxies are also considerably slower than HMC. This is driven by the iterative nature of the NUTS algorithm. As a NUTS chain proceeds, an adaptive step size is set internally through an iterative algorithm so that the sampler avoids taking a ``U-turn" in parameter space (see Appendix~\ref{app:SamplerDetails} for further discussion). While this can help an individual NUTS chain converge more quickly, an implication of this feature is that some chains require much more computation than others. For the case of a single chain, we find that the runtime for HMC and NUTS is comparable, since the computational cost is dominated by the initial transfer of the data across the PCI bus. When running thousands of NUTS chains in parallel, the total runtime for the batch ends up being dominated by the slowest chains, resulting in poor scaling with $N_{\rm chains}$ relative to HMC.

%%%%%%%%%%%%%%%%%%%%%%%%%%%%%%%
%%%%%%%%%%%%%%%%%%%%%%%%%%%%%%%

%%%%%%%%%%%%%%%%%%%%%%%%%%%%%%%%%%%%%%%%%%%%%%%%%%
\section{Conclusion}
\label{sec_conclusion}

Stellar Population Synthesis (SPS) is the leading theoretical framework that connects the physical properties of a galaxy to observations of its spectrum and photometry. Bayesian techniques such as MCMC are widely used in the field to derive confidence intervals on galaxy properties, and the computational demands of conventional implementations of these methods are steep, particularly for applications involving large galaxy samples. In this paper, we have developed an SPS parameter inference pipeline based on DSPS, an SPS library written in JAX, allowing us to explore the potential benefits of GPU-accelerated predictions, as well as alternative algorithms for estimating galaxy posteriors that leverage gradient information such as Hamiltonian Monte Carlo (HMC). We have used our pipeline to revisit the computational demands of Bayesian inference of SPS parameters, generally finding orders-of-magnitude speedup beyond traditional techniques. We summarize our primary conclusions below.

\begin{enumerate}
    \item Our pipeline based on HMC and DSPS is able to derive converged Bayesian posteriors on a 12-dimensional SPS model at a rate of approximately $1000$ galaxies per GPU-minute. For comparison, MCMC runtimes of 25 CPU-hours per galaxy are widely quoted in the literature. Our HMC-based posteriors are in excellent agreement with those derived using the \texttt{emcee} implementation of MCMC, as we show in Appendix~\ref{app:SamplerDetails}.
    
    \item The computational benefits of our HMC-based pipeline are especially advantageous when fitting large galaxy samples at once, as this minimizes the one-time cost of transferring data onto the GPU; the practical gains of running a single HMC on a GPU are limited to speedup factors of $\sim2-5$x.
    
    \item We have systematically studied an approximation to the calculation of galaxy photometry that speeds up the computations by factors of 30-50; the approximation is accurate to a level of $0.1\%$ or better, and the posteriors derived under this approximation present no appreciable loss of accuracy. This approximation is not particular to DSPS, and could be employed by any existing SPS library.
\end{enumerate}

As we have shown, leveraging GPU computing architectures and gradient-based inference algorithms results in orders-of-magnitude speedup beyond conventional approaches to deriving Bayesian posteriors on SPS parameters. Nonetheless, SPS inference for large galaxy samples remains computationally expensive. Even with the gains reported here, estimating SPS posteriors for $10^9$ galaxies observed with LSST will require approximately $20,000$ GPU-hours. This is especially costly considering that any practical analysis of a new dataset requires running numerous chains per galaxy to test for systematics.

We thus expect that AI-based methods will continue to play an important role in SPS analyses of large galaxy samples. There are numerous such applications for which the pipeline we have presented here could provide useful support. For example, high-accuracy AI emulators of photometry and SEDs require large samples of training data, and DSPS is able to generate such training data in a fraction of the runtime of comparable SPS libraries. Techniques such as amortized Neural Posterior Estimation \citep[e.g., SEDflow,][]{hahn_melchior_2022_accel_bayes_sed} require large samples of posteriors for training, and the computational gains of our GPU-parallel pipeline makes it far more practical to create such data. Recent work in cosmological inference has demonstrated the benefits of likelihood emulation \citep{to_etal23_linna}, and the techniques presented here could provide training sets of both likelihood evaluations as well as {\em gradients} of the likelihood. Recent advances in Simulation Based Inference \citep{zeghal_etal_22_sbi_gradients} have shown considerable improvements in sampling efficiency when leveraging the availability of gradient information. We have developed our GPU-parallel pipeline with these and other applications in mind for future work.

%%%%%%%%%%%%%%%%%%%%%%%%%%%%%%%%%%%%%%%%%%%%%%%%%%

%%%%%%%%%%%%%%%%%%%%%%%%%%%%%%%%%%%%%%%%%%%%%%%%%%

\section*{Acknowledgements}
We thank Matt Becker, Evan Blake, Suchetha Cooray, Joel Leja, and Shun Saito for useful discussions, and Changhoon Hahn for comments on an early draft.

Work done at Argonne was supported under the DOE contract DE-AC02-06CH11357. We gratefully acknowledge use of the Bebop cluster in the Laboratory Computing Resource Center at Argonne National Laboratory. This research was supported in part by the National Science Foundation under Grant No. NSF PHY-1748958. APH and AB acknowledge support from NASA under JPL Contract Task 70-711320, “Maximizing Science Exploitation of Simulated Cosmological Survey Data Across Surveys.”

We thank the developers of {\tt NumPy} \citep{numpy_ndarray}, {\tt SciPy} \citep{scipy}, Jupyter \citep{jupyter}, IPython \citep{ipython}, JAX \citep{jax2018github}, conda-forge \citep{conda_forge_community_2015_4774216}, and Matplotlib \citep{matplotlib} for their extremely useful free software. While writing this paper we made extensive use of the Astrophysics Data Service (ADS) and {\tt arXiv} preprint repository.

\bibliographystyle{aasjournal}
\bibliography{bibliography}

\appendix
\counterwithin{figure}{section}

%%%%%%%%%%%%%%%%%%%%%%%%%%%%%%%
\section{SPS model details}
\label{appendix_TheoreticalModel}

In Section~\ref{sec_TheoreticalModel}, we gave a high-level overview of the Stellar Population Synthesis (SPS) model used in our computational benchmarks. In this appendix, we provide a fuller description of the details of the SPS model used to derive our results. 

For a galaxy with fundamental parameters $\theta_{\rm gal}$, observed at redshift $z$, the SED of its composite stellar population is calculated as the weighted sum of Simple Stellar Population (SSP) templates:
\begin{align}
\label{eq:galSEDdef}
    L_{\rm gal} (\lambda|z,\theta_{\rm gal}) = &M_{\star}\sum_{i,j} L_{\rm SSP} (\lambda|\tau_{\rm age}^i, Z^j) F_{\rm att}(\lambda|\theta_{\rm gal}) \nonumber \\
    &\times P_{\rm SSP}(\tau_{\rm age}^i, Z^j|z,\theta_{\rm gal}) \; .
\end{align}
Equation~\eqref{eq:galSEDdef} is identical to Equation~\eqref{eq_galSEDdef} in Section~\ref{sec_TheoreticalModel}, and is repeated here for convenience. We treat the quantity $P_{\rm SSP}$ as a separable function of age and metallicity,
$$P_{\rm SSP}(\tauage, Z) \equiv P_{\rm SSP}(\tauage) \times P_{\rm SSP}(Z) \;,$$
allowing us to rewrite Equation~\eqref{eq:galSEDdef} as follows:
\begin{align}
\label{eq_galSED_decomposed}
    &L_{\rm gal} (\lambda|z,\theta_{\rm gal}) = M_{\star}(z) \sum_{i,j} L_{\rm SSP} (\lambda|\tau_{\rm age}^i, Z^j) F_{\rm att}(\lambda|\theta_{\rm dust}) \nonumber \\
    &\qquad \times P_{\rm SFH}(\tau_{\rm age}^i|z,\theta_{\rm SFH}) P_{\rm met}(Z^j|z,\theta_{\rm met}) \; .
\end{align}
In Equation~\eqref{eq_galSED_decomposed}, we have decomposed the collection of model parameters $\theta_{\rm gal}\equiv\{\theta_{\rm SFH}, \theta_{\rm met}, \theta_{\rm dust}\}.$ In the subsections below, we discuss each of these three modeling ingredients in turn.

%%%%%%%%%%%%%%%%%%%%%%%%%%%%%%%%%%%%%%%%%%%%%%%%%%
\subsection{Star formation history}
\label{sec_SFHModel}

In Equation~\eqref{eq_galSED_decomposed}, the composite SED is multiplied by an overall normalization factor, $M_\star (z|\theta_{\rm gal}),$ which refers to the total stellar mass formed up until the observed redshift $z$. This normalization factor is calculable via an integral of the star formation history:
\begin{equation}
\label{eq:totalMstarFormed}
    M_\star (z|\theta_{\rm gal}) \equiv \int _0^{t_{\rm obs}} dt' \; \dot{M}_\star(t'|\theta_{\rm SFH}) \; ,
\end{equation}
where $t_{\rm obs}$ is the age of the universe at the observed redshift. We decompose the total star formation history of a galaxy into two components: a \textit{smooth} SFH which typically creates the vast majority of stellar mass, and a \textit{bursty} SFH that accounts for the galaxy's stellar mass formed during a recent burst of star formation. In what follows, we describe how we model each of these components.

For the smooth SFH model, we assume a double power-law relation of the form:
\begin{align}\label{eq:SFRsmooth}
    \dot{M}_\star^{\rm smooth}(t|\theta_{\rm SFH}) = M_\star^{\rm formed} \left[ \left( \frac{t}{\tau} \right)^\alpha + \left( \frac{t}{\tau} \right)^{-\beta} \right]^{-1}
\end{align}
where $M_\star^{\rm formed}$ is a normalization factor, $\alpha$ and $\beta$ are the two power-law indices describing the falling and rising slopes, respectively, while $\tau$ is related to the turnover time from one power-law to the other. We choose to use the double power-law due to its flexibility and because it has been shown to provide good fits to simulations \citep{Pacifici_2016,Diemer_2017,Carnall_2018,Mosleh2025}. The parameters controlling the smooth SHF relation are then $\theta_{\rm SFR}^{\rm smooth}=(\log M_\star^{\rm formed}, \tau, \alpha, \beta)$. The stellar age weights, $P_{\rm SFH}^{\rm smooth}$, are computed as the relative contribution to the total stellar mass $M_\star$ from stars in bin $i$ of stellar ages of range $\Delta \tau_{\rm age}^i$:
\begin{align}\label{eq:smoothPsfh}
    &P_{\rm SFH}^{\rm smooth}(\tau_{\rm age}^i|z,\theta_{\rm SFH}) = \frac{1}{M_{\star}(z|\theta_{\rm gal})} \nonumber \\
    &\qquad \times \int_{t_{\rm lo}^i}^{t_{\rm hi}^i} dt' \; \dot{M}_\star^{\rm smooth}(t'|\theta_{\rm SFH})
\end{align}
where the integration limits are $t_{{\rm lo} / {\rm hi}}^i = t' \pm \Delta \tau_{\rm age}^i/2$, and using Equation~\eqref{eq:totalMstarFormed}.

In addition to the stellar population formed during ``smooth" star formation, we separately model the contribution to the stellar population from a recent burst. A common approach to capturing the effect of bursty star formation is to use the power spectral density (PSD) to characterize short-timescale fluctuations \citep[e.g.,][]{Caplar2019_PSD,Wang2020_PSD_I,Wang2020_PSD_II,sun_etal24_bursty_sfh_psd_fire}. In our SFH model, we instead formally decompose $P_{\rm SFH}(\tau)$ as follows:
\begin{align}
\label{eq_composite_age_weights}
    P_{\rm SFH}(\tau_{\rm age}) =& (1-F_{\rm burst})\cdot P_{\rm SFH}^{\rm smooth}(\tau_{\rm age}) \nonumber \\ 
    &+ F_{\rm burst}\cdot P_{\rm SFH}^{\rm bursty}(\tau_{\rm age}) \; .
\end{align}
Rather than calculating $P_{\rm SFH}^{\rm bursty}(\tau_{\rm age})$ via an integral of the SFH, we instead directly parametrize the PDF in terms of $\mathcal{T}(x),$ a triweight approximation to a Gaussian:
\begin{align}
\label{eq_triweight}
    \mathcal{T}(x)\equiv\left\{ 
        \begin{tabular}{ll}
            $(35/96) \left[ 1 - (x/3)^2 \right]^3$, & $\|x\| \leq 3$ \\
            $0$, & $\|x\| > 3$
        \end{tabular}
    \right. \; .
\end{align}
The triweight function is essentially a clipped Gaussian, such that $\mathcal{T}(x)$ vanishes beyond the $3\sigma$ region, with vanishing derivative at the clip. The correspondence between $\mathcal{T}(x)$ and $P_{\rm SFH}^{\rm bursty}(\tau_{\rm age})$ is defined by $x = (\log \tau_{\rm age} - \mu)/\sigma$, $\mu = \log T_{\rm peak}$ and $\sigma = (\log T_{\rm max} - \log T_{\rm peak})/3$. In this expression, $T_{\rm peak}$ corresponds to the peak age of the bursting population, and $T_{\rm max}$ controls the total duration of the burst. Thus, the parameters that control the bursting population are $\theta_{\rm SFH}^{\rm burst}=\{F_{\rm burst}, T_{\rm peak}, T_{\rm max}\},$ and the complete set of parameters controlling star formation history is $\theta_{\rm SFH}=\{\theta_{\rm SFH}^{\rm smooth}, \theta_{\rm SFH}^{\rm burst}\}.$

\begin{figure}
\centering
\includegraphics[width=\columnwidth]{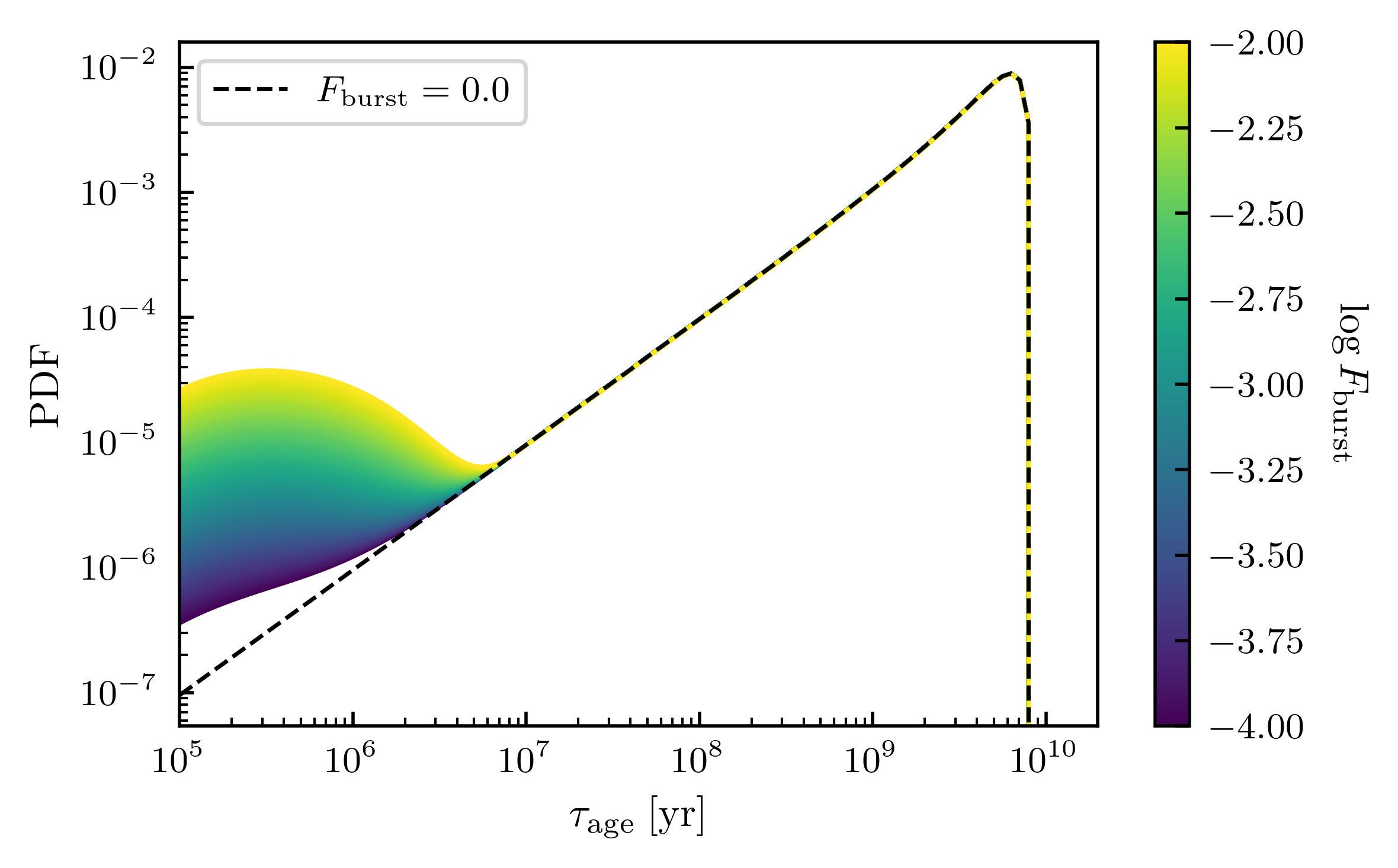}
\caption{\label{appfig:FburstModel} 
Stellar age PDF for a generated galaxy without burstiness ($F_{\rm burst}=0$, dashed line), compared to models with burstiness (solid lines), for a range of $F_{\rm burst}$ parameter values, color-coded as shown in the colorbar.}
\end{figure}

\begin{figure}
\centering
\includegraphics[width=\columnwidth]{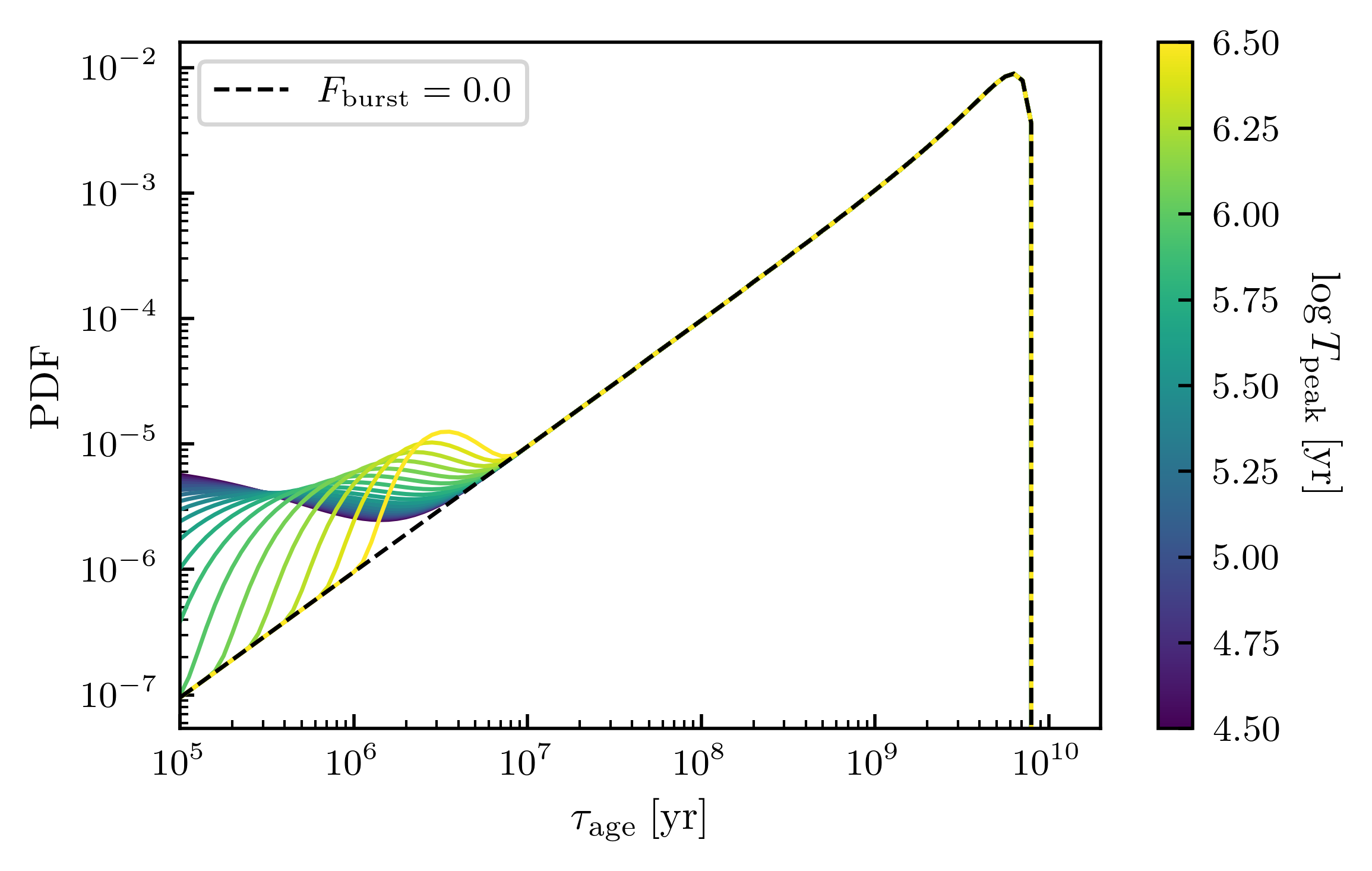}
\caption{\label{appfig:TpeakModel} 
Similar to Figure~\ref{appfig:FburstModel}, but for burstiness models with different $T_{\rm peak}$ parameter values, color-coded as shown in the colorbar.}
\end{figure}

In Figures~\ref{appfig:FburstModel} and \ref{appfig:TpeakModel}, we show how two of the burstiness model parameters, respectively $F_{\rm burst}$ and $T_{\rm peak}$, modify the PDF of stellar ages in an example galaxy. The former, $F_{\rm burst}$, acts as an overall normalization for the bursty part of the model; larger values of $F_{\rm burst}$ correspond to galaxies with a larger total mass formed in a recent burst. The parameter $T_{\rm peak}$ controls the shape of the PDF by effectively shifting the median age of the bursty population; smaller values of $T_{\rm peak}$ correspond to bursts that took place more recently. The effect of the $T_{\rm max}$ parameter (not plotted here) controls the shape of the bursty PDF by modifying its width; larger values of $T_{\rm max}$ correspond to wider PDFs and thus to bursts with a longer duration.

%%%%%%%%%%%%%%%%%%%%%%%%%%%%%%%
%%%%%%%%%%%%%%%%%%%%%%%%%%%%%%%

%%%%%%%%%%%%%%%%%%%%%%%%%%%%%%%
%%%%%%%%%%%%%%%%%%%%%%%%%%%%%%%

%%%%%%%%%%%%%%%%%%%%%%%%%%%%%%%%%%%%%%%%%%%%%%%%%%
\subsection{Metallicity}
\label{sec_MetallicityModel}

The composite SED of a galaxy depends on the full metallicity distribution function (MDF), $P_{\rm SSP}(Z).$ However, galaxy photometry alone has limited constraining power on metallicity, and so it is common for SPS analyses to make the simplifying assumption that this PDF is a delta function centered at some $Z$ that is treated as a free parameter in the fit \citep[e.g.,][]{leja_etal_2020_smf_census}. In this work, we assume a lognormal form for the MDF, so that our model for metallicity includes two parameters, $\theta_{\rm met}=\{\log Z, \sigma_{\log Z}\}.$ The DSPS library includes alternative parametric forms for the MDF besides lognormal, including variations in which stellar age is correlated with metallicity, but variations on the assumed form of the MDF produce no appreciable change in the computational demands of the predictions, and so for purposes of this paper we limit our study to the lognormal form.

%%%%%%%%%%%%%%%%%%%%%%%%%%%%%%%%%%%%%%%%%%%%%%%%%%
\subsection{Dust attenuation}
\label{sec:DustModel}

Due to the presence of dust, some of the light emitted by the composite stellar population in the galaxy is obscured. The reduction of emission due to this effect is modeled by the \textit{dust attenuation curve} $A(\lambda)$, which enters the definition of the function $F_{\rm att}(\lambda) \equiv 10^{-0.4 A(\lambda)}$ by which we multiply the SED, for example in Equation~\eqref{eq_galSEDdef}.

To model the attenuation curve, we use the same functional form adopted in \cite{Salim2018Dust}:
\begin{align}\label{eq:DustAtt}
    A(\lambda) = \left( \frac{A_V}{4.05} \right) \left[ k_0(\lambda) \left( \frac{\lambda}{\lambda_V} \right)^\delta + D_\lambda \right] \; ,
\end{align}
where $A_{\rm V}$ is the normalization defined at $\lambda_{\rm V} = 5500\AA$, $k_0(\lambda)$ is the Calzetti function \citep{calzetti00}, and $D_\lambda$ is the standard Lorentzian-like Drude profile capturing the UV dust bump \citep{fitzpatrick_massa_1986}.

In addition to parameters $A_{\rm V}$ and $\delta$, we include an additional parameter $F_{\rm floor}$ that accounts for the fraction of sightlines through the galaxy that are unobscured by dust:
\beq
F_{\rm att}(\lambda) = F_{\rm floor} + (1-F_{\rm floor}) 10^{-0.4 A(\lambda)} \; .
\eeq
In \citet{Lower2022Dust}, the parameter $F_{\rm floor}$ was shown to improve the description of the effect of dust attenuation on mock SEDs in the Simba simulation \citep{dave_etal19_simba}, and in \citet{hahn_melchior_2025_clumpy_dust} it was demonstrated that $F_{\rm floor}$ likely plays an important role in mitigating biases in analyses of optical photometry.

%%%%%%%%%%%%%%%%%%%%%%%%%%%%%%%
\section{Approximate photometry model}
\label{appendix_ApproxPhotoModel}
In this Appendix, we present further results assessing the accuracy of the approximation in the photometry computation discussed in Section~\ref{sec_Photometry}. We performed two separate accuracy tests: first, we examine the fractional error on the predicted photometry; second, we compare the inferred parameter posteriors with HMC. We discuss these two tests in turn below.

%%%%%%%%%%%%%%%%%%%%%%%%%%%%%%%
\begin{figure*}
\centering
\includegraphics[width=\textwidth]{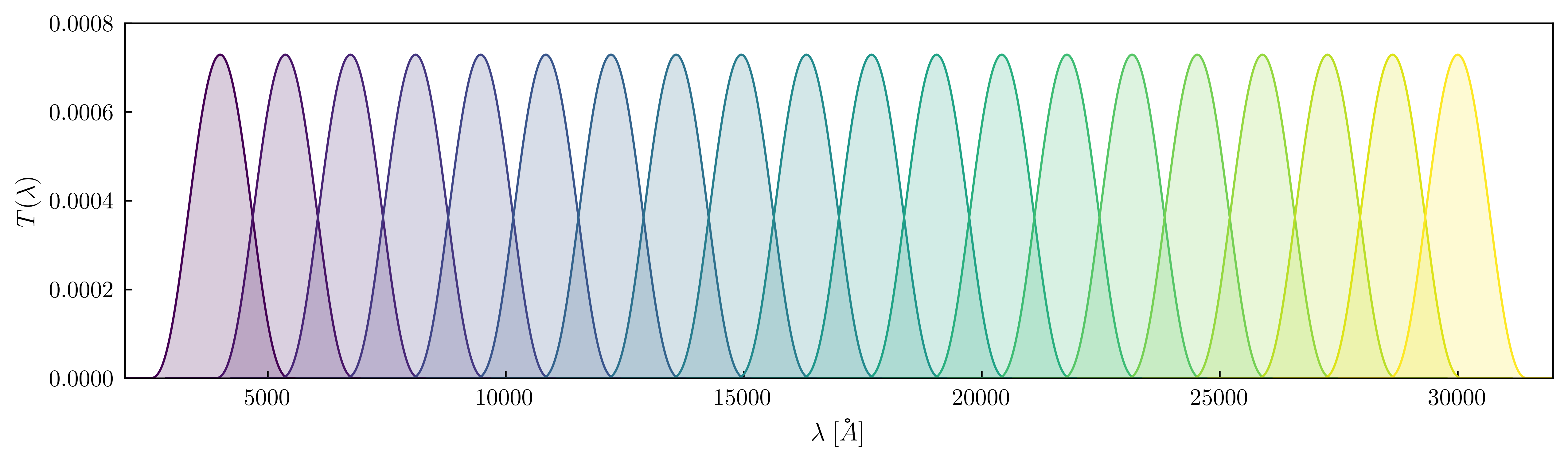}
\caption{\label{appfig:trans_bands} 
Synthetic gaussian transmission bands used to produce synthetic data and during the model fits.}
\end{figure*}
%%%%%%%%%%%%%%%%%%%%%%%%%%%%%%%

\begin{figure}
\centering
\includegraphics[width=\columnwidth]{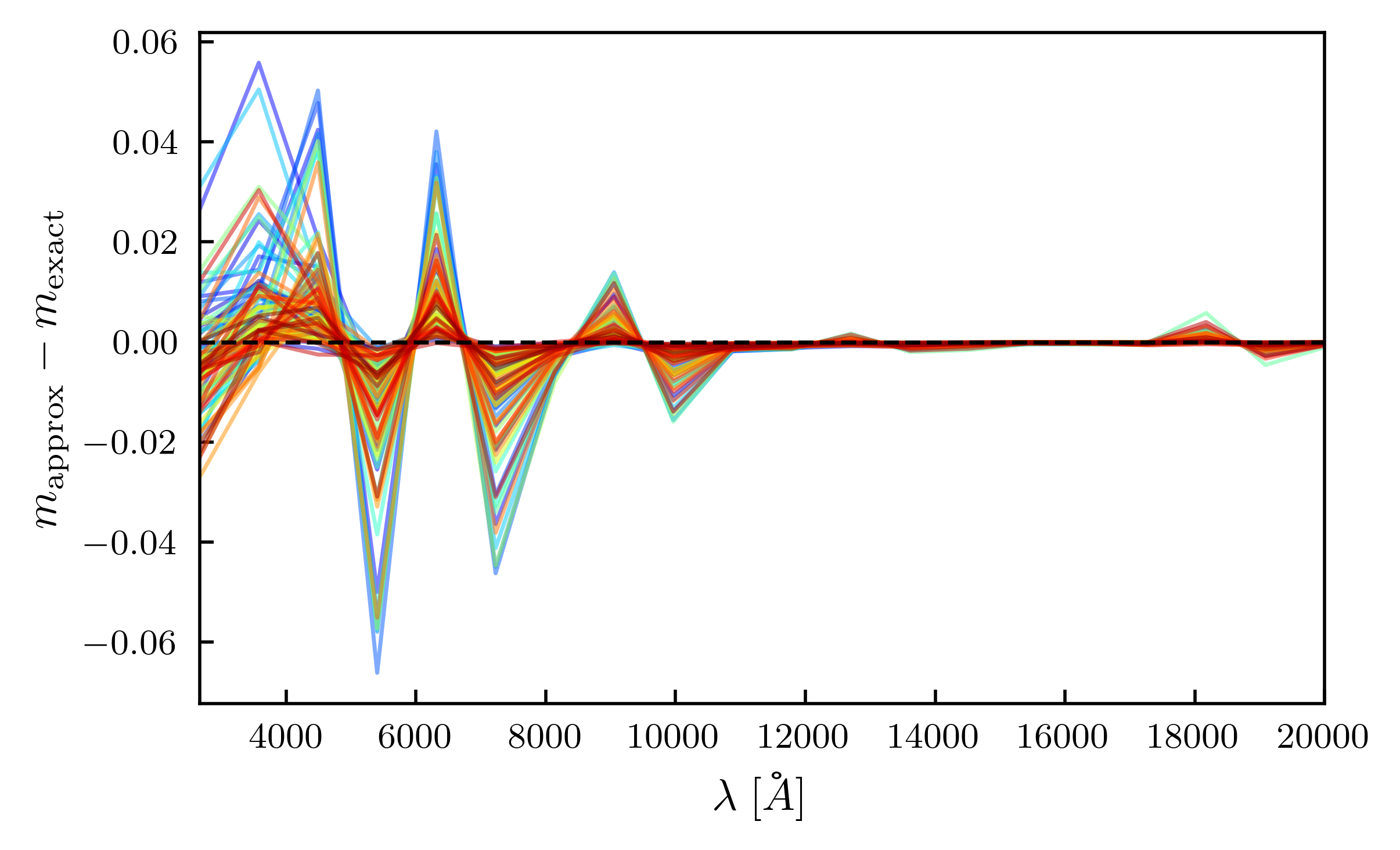}
\caption{\label{appfig:PhotoApproxFracErr} 
Difference between the approximate model for the apparent magnitude, $m_{\rm approx}$, and the exact model, $m_{\rm exact}$. Different color lines represent different choices of $\theta_{\rm gal}$ sampled uniformly from the priors.}
\end{figure}

\begin{figure}
\centering
\includegraphics[width=\columnwidth]{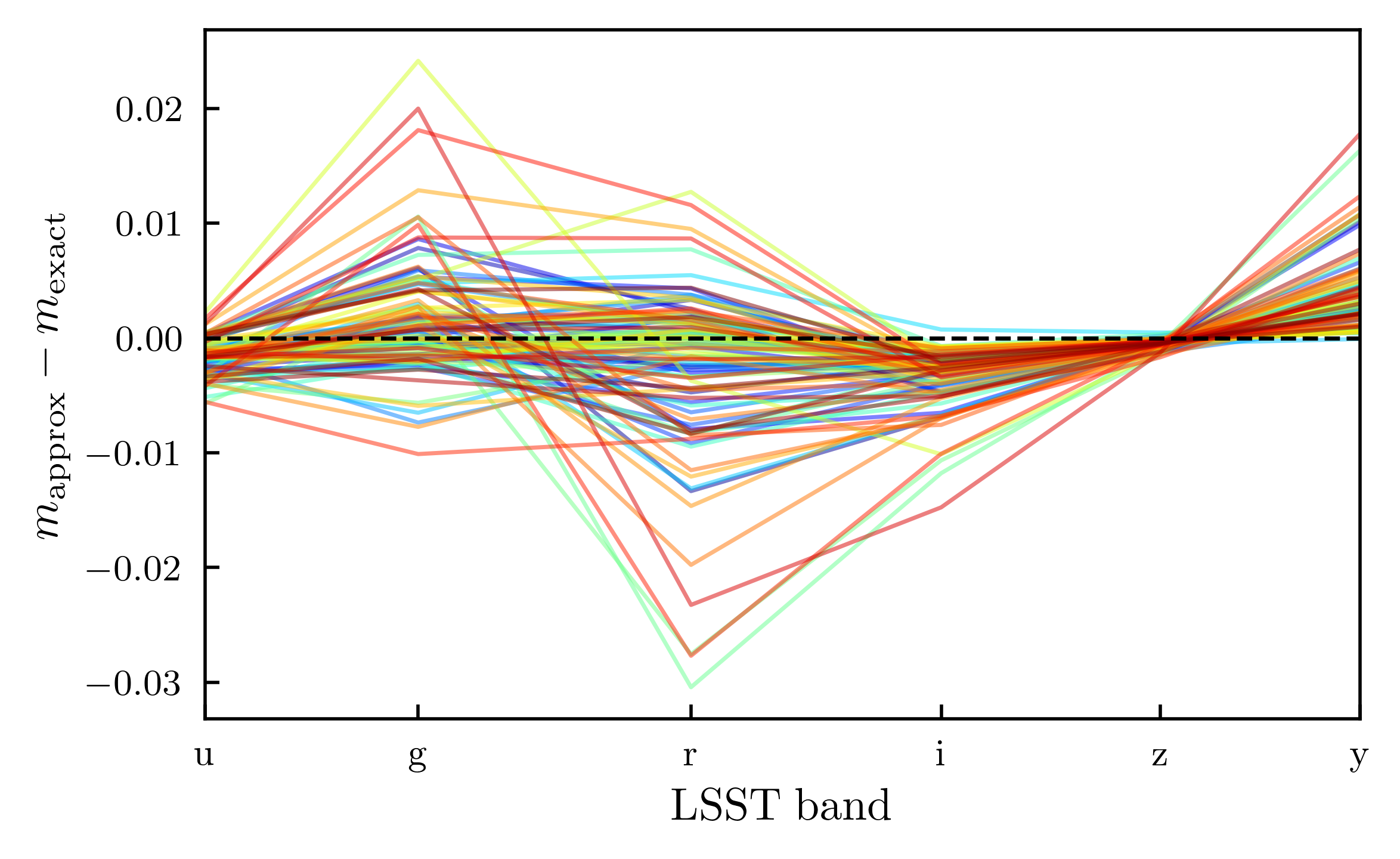}
\caption{\label{appfig:PhotoApproxFracErrLSST} 
Similar to Figure~\ref{appfig:PhotoApproxFracErr} but this time using the 6 LSST transmission bands $(u,g,r,i,z,y)$ for a more realistic test on the accuracy of the approximate photometry compared to the exact model.}
\end{figure}

Figure~\ref{appfig:PhotoApproxFracErr} shows the fractional differences between the apparent magnitude of a galaxy using the approximate versus the exact model. We have produced these model predictions using the gaussian transmission filters shown in Figure~\ref{appfig:trans_bands}. The range of resolution of these filters, defined as the ratio of the mean wavelength to the full width at half maximum (FWHM) of a transmission cure, $\lambda_{\rm mean} / {\rm FWHM}$, varies from $\sim 3.4$ for the left-most band to $\sim 25.5$ for the right-most one in Figure~\ref{appfig:PhotoApproxFracErr}. We perform the accuracy test for a large number of different galaxy properties $\theta_{\rm gal}$, by uniformly sampling from the parameter prior distributions (see Appendix~\ref{appendix:ParameterBounding} for more on the priors). The approximation is accurate to better than $0.06$ magnitude for a wide range of wavelengths. Note, however, that the level of accuracy of the approximation depends on the width of the transmission band: the broader the band, the less accurate this approximation will be. Thus the test we present here should be repeated before applying it to an analysis of a new dataset. As a test case for the applicability of our approximate model to more realistic transmission bands, we performed a test that is similar to the above using the 6 LSST bands $(u,g,r,i,z,y)$. We computed theoretical photometry predictions for various sets of model parameters comparing the exact with the approximate photometry model, in both cases using the LSST bands, and the results are presented in Figure~\ref{appfig:PhotoApproxFracErrLSST}. As we can see, the approximate model works well in this case as well, always resulting in magnitude differences less than $0.03$.

\begin{figure*}
\centering
\includegraphics[width=\textwidth]{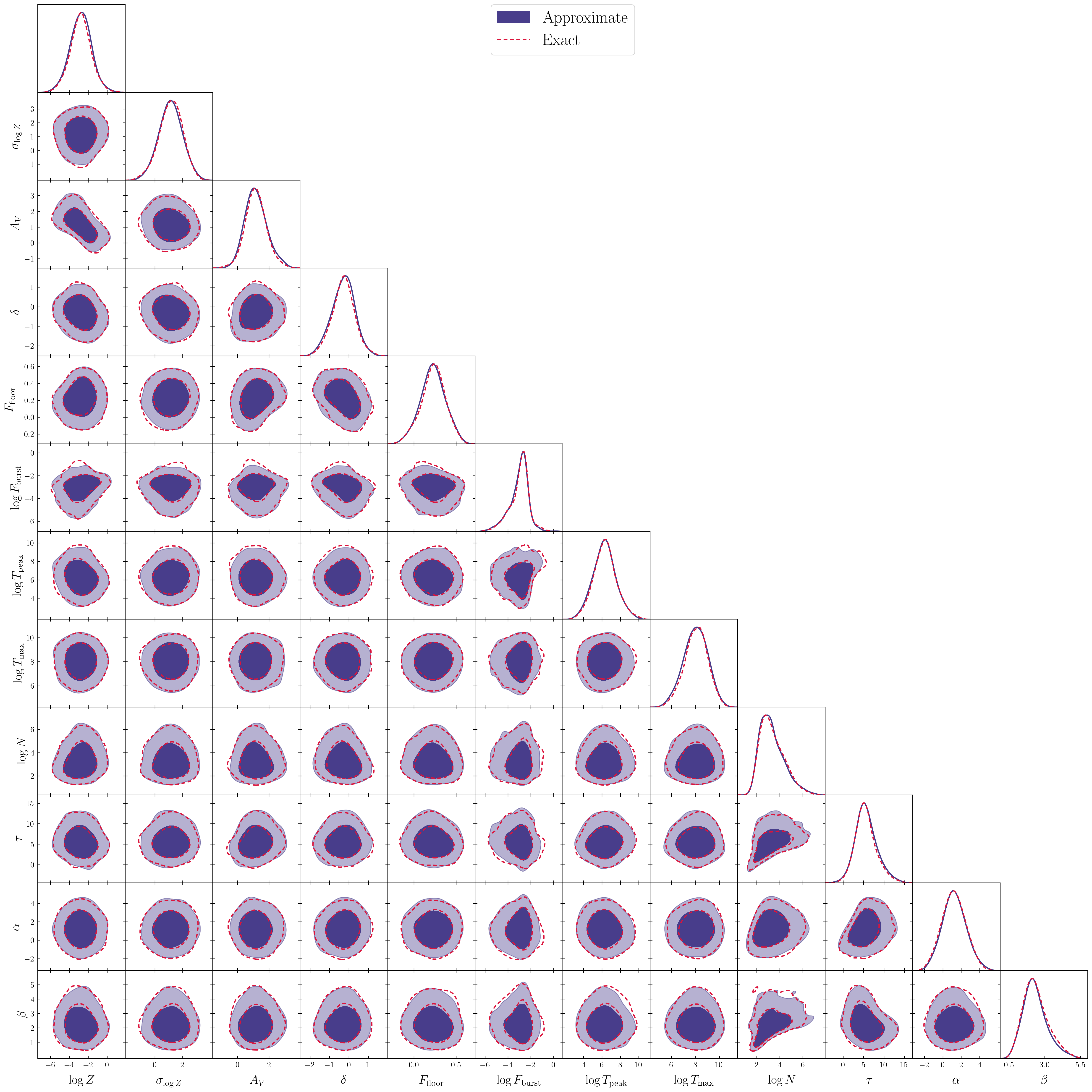}
\caption{\label{appfig:ApproxModelTestAllParams} 
Comparison between running a HMC inference using the approximate and exact photometry model, as described in Section~\ref{sec_Photometry}, showing all 12 model parameters.}
\end{figure*}

For the second test, we compare the posteriors from HMC using both the exact and the approximate photometry models when fitting an example galaxy with 12 model parameters. We perform this test using the exact same initial conditions for both fits and the results are shown in Figure~\ref{appfig:ApproxModelTestAllParams} for the full parameter space. We find that the posteriors from the approximate and exact photometry calculation are nearly indistinguishable. The approximation also speeds up the HMC calculation by greater than $30\times$: for a converged chain with $3000$ steps, using the exact photometry takes 11 hours, while the runtime is 21 minutes when using the approximation.

%%%%%%%%%%%%%%%%%%%%%%%%%%%%%%%
%%%%%%%%%%%%%%%%%%%%%%%%%%%%%%%

%%%%%%%%%%%%%%%%%%%%%%%%%%%%%%%
%%%%%%%%%%%%%%%%%%%%%%%%%%%%%%%

%%%%%%%%%%%%%%%%%%%%%%%%%%%%%%%
%%%%%%%%%%%%%%%%%%%%%%%%%%%%%%%

%%%%%%%%%%%%%%%%%%%%%%%%%%%%%%%
\section{Parameter bounding and prior distributions}
\label{appendix:ParameterBounding}

In this appendix, we discuss how we construct the prior distributions in the Bayesian sampling algorithms we used in the main body of the text. Measurements of galaxy photometry provide a fundamentally limited amount of information about the fundamental properties of the galaxy, such as its star formation history \citep{ocvirk_etal06}, and even in high signal-to-noise measurements the posterior information can be prior-dominated \citep{Carnall_2019_SFH,leja_etal19_how_to_measure2}. For the purposes of this paper, we have opted for simple parametric families of PDFs for the prior, since our aims are primarily to assess matters of computational performance of running HMCs with stellar population synthesis on GPUs. As a much more flexible alternative, generative neural networks such as PopSED \citep{li_etal24_popsed} or pop-cosmos \citep{alsing_etal24_pop_cosmos} can capture  complexity that is not possible with simple parametric families; depending on the size and complexity of the network, AI-based priors may come with a non-trivial computational cost and memory footprint when evaluating at scale that we do not consider here. Instead, we focus here on basic techniques for constructing  families of priors while respecting custom bounds on the parameters of a physical model. First, we describe how we construct {\it unbounded} versions of the parameters used in our fits. We then describe how we set up priors on the unbounded parameters, and finally transform the PDFs back into the bounded parameter space.

Consider an SPS parameter, $\theta,$ which on physical grounds should be bounded within some finite range. While running a gradient-based algorithm for either inference or optimization, there is no guarantee that a step in the direction of the gradient will respect the bounds. Some algorithms are naturally formulated to take such bounds into account \citep[e.g., BFGS,][]{broyden_1970_B_in_BFGS,fletcher_1970_F_in_BFGS,goldfarb_1970_G_in_BFGS,shanno_1970_S_in_BFGS}, while others such as Adam \citep{kingma_ba_adam_2015} are less adaptable. To address this issue, we construct \textit{unbounded} parameters, $\tilde{\theta}$, which are related to $\theta$ through a {\it bounding function}, $\mathcal{F}_{\rm B},$ an invertible bijective transformation:
\beq
\theta&=&\mathcal{F}_{\rm B}(\tilde{\theta}) \; , \nonumber\\
\tilde{\theta}&=&\mathcal{F}^{-1}_{\rm B}({\theta}) \; .\nonumber
\eeq
The unbounded parameters $\tilde{\theta}$ can take on any value on the real line, since the bounding function $\mathcal{F}_{\rm B}$ guarantees that the physical parameters $\theta$ will always remain in bounds. We then carry out the optimization or inference algorithm on the unbounded space, and then transform back into the bounded, physical parameter space to report best-fitting values and constraints.
Probabilistic programming languages such as {\tt Stan}\footnote{\url{https://mc-stan.org/docs/2_18/stan-users-guide/}} construct such bounding functions programmatically under the hood; in this paper, we construct custom bounding functions based on the following sigmoid function:
\begin{align}\label{eq:BoundingFunction}
    \theta &= \mathcal{F}_{\rm B}(\tilde{\theta} | p_\theta) \equiv y_{\rm lo}^{\theta} + \frac{y_{\rm hi}^{\theta} - y_{\rm lo}^{\theta}}{1 + \exp[-k^{\theta} (\tilde{\theta} - x_0^{\theta})]} \; .
\end{align}
In Equation~\eqref{eq:BoundingFunction}, the hyper-parameters of the sigmoid function for parameter $\theta$ are $p_\theta = \{ y_{\rm lo}^{\theta}, y_{\rm hi}^{\theta}, k^{\theta}, x_0^{\theta} \}$; these hyper-parameters need to be tuned in advance of running the sampler, and thereafter held fixed; the quantities $y_{\rm lo}^{\theta}$ and $y_{\rm hi}^{\theta}$ define the bounds on $\theta,$ while $k^{\theta}$ and $x_0^{\theta}$ should be tuned so that the behavior of $\mathcal{F}_{\rm B}$ is approximately linear in the range expected for the best-fit $\theta.$ More complex bounding functions than these can naturally arise when imposing constraints on a model; for example, the Diffstar parameterization for galaxy SFH \citep{diffstar_alarcon_etal23} uses bounding functions that capture constraints on the parameter based on the physics of galaxy formation; bounding functions are typically lightweight, fast-evaluating functions that contribute negligibly to the computational demands of the Bayesian inference.

Once the bounding function $\mathcal{F}_{\rm B}$ is defined, it is straightforward to use it to construct families of priors in the space of unbounded parameters, $\tilde{\theta}$. We start with a simple assumption of approximately Gaussian priors on $\theta,$ draw $N$ samples, $\{{\theta}_{\rm i}\},$ and use $\mathcal{F}^{-1}_{\rm B}(\theta_{\rm i})$ to map into the unbounded space, $\{\tilde{\theta}_{\rm i}\}$. We parameterize $P(\tilde{\theta})$ using a skew-normal distribution:
    \begin{align} \label{eq:SkewNormPrior}
        \mathcal{P}(\tilde{\theta}) = \frac{2}{\omega} \mathcal{N}\left( \frac{\tilde{\theta}-\xi}{\omega} \right) \Phi \left( s \left( \frac{\tilde{\theta}-\xi}{\omega} \right) \right) \; ,
    \end{align}
where $$ \mathcal{N}(x) = \frac{1}{\sqrt{2\pi}} e^{-x^2/2} $$ is the normal distribution. For each parameter $\theta,$ we fit the functional form in Equation~\eqref{eq:SkewNormPrior} to the distribution of samples $\{\tilde{\theta}_{\rm i}\}$ to determine the set of skew-normal parameters $(\xi_{\theta}, \omega_{\theta}, s_{\theta})$. We thus use the best-fitting skew-normal distributions to define the priors on $\tilde{\theta}$, so that the priors on $\theta$ are defined implicitly via $\mathcal{F}_{\rm B}.$

%%%%%%%%%%%%%%%%%%%%%%%%%%%%%%%
\section{Sampler details}\label{app:SamplerDetails}

In this Appendix, we compare the gradient-based samplers we have utilized in this work with each other, and discuss advantages and limitations of each. We also briefly discuss how these samplers compare to traditional MCMC algorithms.

%%%%%%%%%%%%%%%%%%%%%%%%%%%%%%%
\subsection{Simulated galaxy sample}

The galaxy sample we use in this work to fit our model to is simulated. In particular, we use the SPS model that is described in Appendix~\ref{appendix_TheoreticalModel} to both produce the simulated dataset and to make model predictions while running our inference pipeline. The simulated sample consists of 20000 galaxies, each of which is described by a set of 12 parameters, sampled from the prior distributions, to characterize the galaxy's star-formation history, mass-metallicity relation, dust attenuation, and burstiness properties. In correspondence with these parameter sets, we generate mock photometry for each galaxy using the mathematical framework we describe in Section~\ref{sec_TheoreticalModel} and Appendix~\ref{appendix_TheoreticalModel} with the transmission bands presented in Figure~\ref{appfig:trans_bands}. Thus, each galaxy is also described by a 20-point photometry vector. In addition, we design a diagonal gaussian covariance matrix to accompany this data vector, defined as $\mathcal{C} = 0.2 \times \mathbb{I}_{20}$, where $\mathbb{I}_{20}$ is the $20\times 20$ identity matrix. The gaussian width $\sigma_{\rm phot}=0.2$ of this covariance is chosen as a simple but reasonable value of the uncertainty on the photometry of each galaxy.

%%%%%%%%%%%%%%%%%%%%%%%%%%%%%%%
\subsection{Gradient-based sampler details}

The two samplers we consider that utilize gradients are Hamiltonian Monte Carlo \citep[HMC,][]{duan_etal87_hmc,neal_2011_hmc_chapter} and the No-U-Turn Sampler \citep[NUTS,][]{hoffman_gelman_2014_nuts}. HMC simulates a physical system with a potential and a kinetic term that explores the parameter space using the dynamics of the system. In particular, the \textit{leapfrog integrator} is utilized by the algorithm for the system to move from one position to the next. The main benefit from using HMC over MCMC derives from the usage of gradients, which are incorporated into the kinetic term; having access to gradient information allows the sampler to construct trajectories through parameter space that are informed by the structure of the likelihood surface, reducing the noisy random-walk behavior of conventional MCMC samplers.

The \textit{No-U-Turn Sampler} (NUTS) is an adaptive variant of HMC that includes a programmatically-determined stopping criterion. In standard HMC, the number of leapfrog steps is set in advance, which can lead to suboptimal sampling efficiency. NUTS addresses this limitation by dynamically detecting when the trajectory begins to reverse direction -- a ``U-turn" -- and using this as an automated stopping condition. This feature improves the sampling efficiency of NUTS without the need to manually tune the trajectory length.

In this work, when deriving posteriors on our SPS model parameters constrained by photometry, we generally find that for a typical individual chain, the convergence rate of NUTS is  considerably faster than that of traditional HMC: it is challenging to determine in advance the appropriate number of leapfrog integration steps to achieve convergence for HMC, whereas NUTS takes care of this task adaptively while running. However, for a fixed number of steps, NUTS is generally slower than HMC, as the task of dynamically determining the number of integration steps comes with additional computational cost. Furthermore, the advantage of the adaptive stopping criterion of NUTS becomes a disadvantage when running large numbers of fits in parallel on a GPU. Since each galaxy fit has its own unique stopping condition, running tens of thousands of fits in parallel typically ends up being as slow as the slowest chain to run. As shown in Section~\ref{section_FittingPipeline}, this leads to faster time-to-solution of HMC over NUTS when deriving posteriors for large galaxy samples with different chains running on independent GPU threads.

%%%%%%%%%%%%%%%%%%%%%%%%%%%%%%%
\subsection{Sampler comparison}\label{app:SamplerComparison}
\begin{figure*}
\centering
\includegraphics[width=\textwidth]{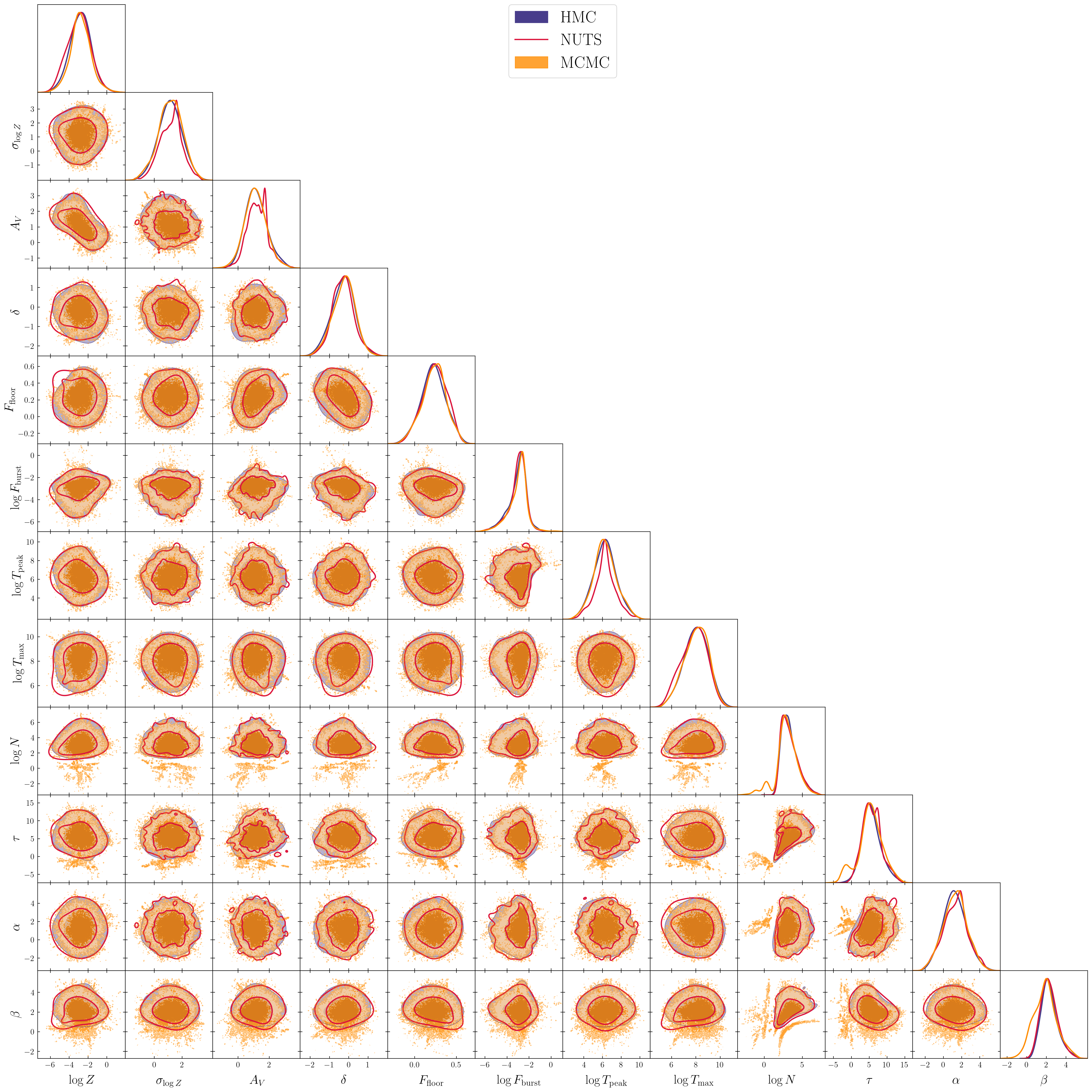}
\caption{\label{appfig:HMCvsMCMCvsNUTS} 
Comparison between HMC, NUTS and MCMC, showing all 12 model parameters.}
\end{figure*}

We have carried out a simple comparison of the results of each sampler to ensure that we derive comparable posteriors when adopting different sampling algorithms. We have run a fit to the same target data using both HMC and NUTS, as well as {\tt emcee} \citep{EMCEEpaper}, a widely used MCMC implementation. In Figure~\ref{appfig:HMCvsMCMCvsNUTS} we present the parameter posteriors inferred from running HMC, NUTS and MCMC on the same target data. All chains have converged based on the $\hat{R}$ statistic values for all 12 model parameters. The runtimes for HMC, NUTS and MCMC are, respectively, $22$, $23$ and $29$ minutes. The MCMC sampler ran for $30,000$ steps with $32$ walkers, while HMC and NUTS ran for $3,000$ steps each, with additional $100$ steps of adaptation to tune the sampling parameters. The three samplers produce very similar contours. The contours of HMC and NUTS are visibly smoother than {\tt emcee} due to the stochastic nature of MCMC sampling. The comparison to {\tt emcee} gives non-trivial validation that our gradient-based posteriors are comparable to MCMC-based methods that are more widely used in the field of SPS; the comparison of HMC to NUTS also demonstrates that the HMC hyper-parameters that control the leapfrog integrator were tuned properly by the adaptation phase.

%%%%%%%%%%%%%%%%%%%%%%%%%%%%%%%
%%%%%%%%%%%%%%%%%%%%%%%%%%%%%%%

\end{document}